\documentclass[12pt]{article}

\usepackage{amsmath,amsfonts,amssymb,latexsym}

\numberwithin{equation}{section}
\def\be{\begin{equation}}
\def\ee{\end{equation}}
\def\bq{\begin{eqnarray}}
\def\eq{\end{eqnarray}}
\def\beq{\begin{eqnarray*}}
\def\eeq{\end{eqnarray*}}

\def\a{\alpha}

\def\g{\gamma}
\def\d{\delta}

\def\pa{\partial}

\begin{document}
\begin{titlepage}
\begin{flushright}
{\tt CERN-PH-TH/2014-097}\\
\end{flushright}

\vspace{0.3cm}

\begin{center}
{\Huge Enveloping branes and braneworld singularities}

\vspace{1cm}
{\large Ignatios Antoniadis$^{1,*,3}$, Spiros Cotsakis$^{1,\dagger,4}$, Ifigeneia Klaoudatou$^{2,\ddagger}$}\\

\vspace{0.5cm}

$^1$ {\normalsize {\em Department of Physics, CERN - Theory Division}}\\
{\normalsize {\em CH--1211 Geneva 23, Switzerland}}\\

\vspace{2mm}

$^2$ {\normalsize {\em Research Group of Geometry, Dynamical Systems
and Cosmology}}\\ {\normalsize {\em Department of Information and
Communication Systems Engineering}}\\ {\normalsize {\em University
of the Aegean,}} {\normalsize {\em Karlovassi 83 200, Samos,
Greece}}\\
\vspace{2mm} {\normalsize {\em E-mails:}
$^*$\texttt{ignatios.antoniadis@cern.ch},
$^\dagger$\texttt{skot@aegean.gr},
$^\ddagger$\texttt{iklaoud@aegean.gr}}
\end{center}

\vspace{0.5cm}

\begin{abstract}
\noindent
The existence of envelopes is studied for systems of differential equations in connection with
the method of asymptotic splittings which allows to determine the singularity structure of the
solutions. The result is applied to braneworlds consisting of a 3-brane in a five-dimensional
bulk, in the presence of an analog of a bulk perfect fluid parametrizing a generic class of bulk
matter. We find that all flat brane solutions suffer from a finite distance singularity contrary
to previous claims. We then study the possibility of avoiding finite distance singularities by
cutting the bulk and gluing regular solutions at the position of the brane. Further imposing
physical conditions such as finite Planck mass on the brane and positive energy conditions
on the bulk fluid, excludes however this possibility, as well.
\end{abstract}

\begin{center}
{\line(5,0){280}}
\end{center}

$^3${\small On leave from {\em CPHT (UMR CNRS 7644) Ecole
Polytechnique, F-91128 Palaiseau, France.}}

$^4${\small On leave from the University of the Aegean, 83200 Samos, Greece.}

\end{titlepage}
\section{Introduction}
In a previous work \cite{ack}, we studied and classified the singularity structure and the
corresponding asymptotic behavior of a 3-brane in a five-dimensional (5d) bulk, in the
presence of an analog of a bulk perfect fluid, using the so-called method of asymptotic
splittings \cite{skot}, \cite{goriely}. We assumed that the bulk fluid satisfies an equation
of state $p=\gamma\rho$, with a constant parameter $\gamma$, while the `pressure' $p$
and the `density' $\rho$ are functions of the fifth coordinate $Y$. We found a surprising
result that the flat brane solution does not suffer from a finite distance singularity in the
region $-1<\gamma\le -1/2$, opening the possibility of the self-tuning mechanism for the
physical cosmological constant.

More precisely, this conclusion was reached as follows: for $\gamma$ outside this region,
the flat brane solution found asymptotically was general, i.e. with maximum number of
arbitrary constants, and had a singularity at finite distance from the brane position. For
$\gamma$ in the above region on the other hand, the singular flat brane solution had less
constants and was thus considered to be particular with the general solution assumed  regular having
a singularity at infinity. On the other hand, as we will see later, it is possible to find the
general solution for the flat brane explicitly and is singular for any value of $\gamma$,
which created a puzzle for the meaning of the asymptotic solutions.

The solution of the puzzle is the existence of envelopes in a system of differential equations
\cite{goursat}, \cite{goursat2}, together with the correct interpretation of those asymptotic solutions which show a blow up at infinity. Envelopes amount to solutions with a smaller number of arbitrary
constants, and are configurations  of a special nature that the asymptotic method can also pick. In the first part of this work, we study the
existence and properties of envelopes and discuss their consequences in the singularity
structure analysis we made in \cite{ack}. It turns out that the solution we found with the
method of asymptotic splittings in the above mentioned region of $\gamma$ is not
particular but a general envelope.

In the second part of this work, we study the possibility of avoiding finite distance
singularities by cutting the bulk and gluing two non-singular branches of solutions at the
position of the brane \cite{forste}. We find that this is indeed possible, while the condition
of finite four-dimensional (4d) Planck mass restricts the region of $\gamma$ to
$-2<\gamma<-1$. We then study the possibility of having physical systems with such an
equation of state by analyzing the energy conditions for a bulk perfect fluid and we find that
this region is excluded.

The plan of this paper is the following: In Section 2, first we describe briefly the concept
of envelopes; in Subsection 2.1 we give some simple examples and analyze their effect in general, while
in Subsection 2.2 we set-up the dynamics of our model and analyze the nature of the
envelopes that it exhibits. We then compare this result with the
asymptotic behaviors that we found for the same model in \cite{ack}. It follows that
there always exist finite-distance singularities for all values of the parameter of the fluid,
$\gamma$. Since there is no way of avoiding these singularities when the bulk space is
considered as an indivisible entity, in Section 3, we
exploit the presence of the brane that introduces a natural symmetry in the bulk and
explore the possibility of avoiding finite singularities by cutting the bulk space and
matching the solutions that are regular (i.e. they exhibit no finite singularities). In
Subsection 3.1, we examine which range of $\gamma$ gives a finite four-dimensional
Planck mass, while in Subsection 3.2, we investigate whether this range of $\gamma$
satisfies physical constraints such as the weak and strong energy conditions. In Section 4,
we conclude and comment on questions that remain open within the framework of the class
of models considered in this paper. In Appendix A, we analyze the envelopes that exist
in the case of a flat or curved brane in a perfect fluid bulk for the various values
of $\gamma$. Lastly, in Appendix B, we derive in detail the forms of the weak
(Subsection B.1) and strong (Subsection B.2) energy conditions.
\section{Asymptotic behavior and existence of envelopes}
In this Section, we review the effect of the existence of envelopes on differential equations
in general but also for the system of differential equations that describes the braneworld
model we studied in \cite{ack}.

First, we describe briefly the basic concepts and terminology that we use, starting with the
definition of an envelope. Consider a one-parameter family of curves described by the
equation
\be
\label{fam curves}
F(x,y,c)=0,
\ee
where $c$ is the arbitrary parameter. 
An \emph{envelope}, $\mathcal{E}$, is a fixed curve that is tangent to all members of this
family of curves at some point. Thus, the slope of a member of the family is the same
with the slope of the envelope at the point of intersection. This condition combined with
the fact that the envelope itself satisfies the equation of the family (\ref{fam curves})
at the point of intersection leads to the following set of equations \cite{goursat}
\be
\label{envelope}
F(x,y,c)=0 \quad \textrm{and} \quad \partial_{c} F=0.
\ee
The equation of the envelope can be derived by elimination of the parameter $c$ in the
above set of equations. It can be shown \cite{goursat} that the resulting equation
includes also the locus of the critical points.

This analysis can be extended to higher dimensional objects. For instance, if instead of a family of curves we have a family of surfaces given by the
equation
\be\label{gen sur}
f(x,y,z;c)=0,
\ee
and there exists a surface $\mathcal{E}$, which is tangent to each member of this family
of surfaces along a curve, then the surface $\mathcal{E}$ is the \emph{enveloping surface}
of (\ref{gen sur}). We may define the enveloping surface by elimination of $c$ in the
equations
\be\label{sfenv}
f(x,y,z;c)=0,\quad\pa_c f=0.
\ee
Then it can be shown that the resulting equation consists of two in general analytically
distinct groups of surfaces, one of which is the envelope of the original surface and the
other is the locus of critical points, that is the set $\nabla f=0$, cf. \cite{goursat2}, Sect. 219.

The method described above may be applied to the general solution of a differential equation,
in order to trace the envelope which is usually a solution that cannot be derived
from the general solution by assigning a particular value to some arbitrary constants. Since this method may introduce other points
than the envelope, a check should be performed as a last step to certify that any curve
found is indeed a solution of the system of differential equations under consideration
\cite{ford}, p.~17-18.
\subsection{Simple examples}
Here we give simple examples of differential equations to explain in general
a novel relation between envelopes and asymptotic behavior traced by the method expounded
in \cite{skot} and \cite{goriely} which we used in our analysis of braneworld singularities in
\cite{ack}.
We shall show that our asymptotic method called `asymptotic splittings' has the further property that if the general solution
has an envelope\footnote{In general, any system of dynamical equations may itself define an envelope (for instance, when one considers a differential system with constraints) irrespective of whether or not the general solution is known. A general theory connecting the method of asymptotic splittings with the existence of envelopes for a general differential dynamical system is at present unknown.}, that is a limiting curve or surface to which all members of the family
become tangents, then it is this enveloping curve that may be picked by the method instead
of the general solution. On physical and geometrical grounds, this is to be expected since the method of asymptotic splittings is concerned with the asymptotic nature of the solutions, and if a dynamical system has an envelope, then its solutions will be asymptotically tangent  to it.

We first consider the equation
\be\label{ex1}
\dot{x}^2-\dot{x}t+x=0,
\ee
which has as a general solution the one-parameter family of curves \cite{tabor}
\be\label{g1}
x=ct-c^2,
\ee
where $c$ is a constant. Following the method of asymptotic splittings, substituting the ansatz
\be\label{db}
x=at^p,\quad a,p\,\,\textrm{constants,}
\ee
which we call a dominant balance,
we find that the equation also admits the following solution,
\be\label{s1}
x=\frac{t^2}{4}.
\ee
(This is also noted in \cite{tabor}, albeit using a different  method.) This solution has no
arbitrary constant (that is it has one less than the general solution) and does not follow from
the general solution (\ref{g1}). But the solution (\ref{s1}) is the envelope of (\ref{g1}),
cf. \cite{tabor}, pp. 333-4, Fig. 8.1. Another way to see the enveloping
property of (\ref{s1}) for the family (\ref{g1}), is to set $F(x,t,c)=ct-c^2-x$, and the envelope
property means the simultaneous validity of the equations
\be\label{env}
F=0,\quad \pa_c F=0.
\ee
Then $\pa_c F=0$ gives $t-2c=0$, or $c=t/2$, and from the first equation we find
that $x=2c^2-c^2$, or $x=t^2/4$.

We therefore conclude that the dominant balance picks the envelope not the general
solution, when the latter has an envelope.

A more interesting example is the equation
\be\label{ex2}
\dot{x}^2+\dot{x}x^2t+x^3=0,
\ee
which has the general solution,
\be\label{g2}
x=\frac{1}{ct-c^2}.
\ee
We substitute the dominant balance (\ref{db}) and find the nontrivial solution $a=4,p=-2$,
that is
\be\label{s2}
x=\frac{4}{t^2},
\ee
and we naturally wonder whether this is the envelope of (\ref{g2}). Such an envelope,
if it exists, has to satisfy the equations (\ref{env}), for the function
\be
F=x-\frac{1}{c(t-c)}.
\ee
Then we calculate
\be
\pa_cF=\frac{-t+2c}{(c(t-c))^2},
\ee
so that $c=t/2$ which putting it back in Eq. (\ref{g2}), leads to $x$ actually being given by
(\ref{s2}).
Thus, we find again that the dominant balance picks the envelope instead of the general
solution. Note the dependence in (\ref{s2}), which is very similar to the asymptotic forms
of the density we found typically in our work \cite{ack}.

The condition to deduce the existence of an envelope from the first order differential
equation itself is that \cite{goursat2}, Sections 71-4, the equation $F(\dot{x},x,t)=0$
has a double root in $\dot{x}$. For example, setting $\varphi=\dot x$ in Eq. (\ref{ex1}),
we get
\be
\varphi^2-\varphi t+x=0,
\ee
and this has a double root provided the discriminant vanishes, that is
\be
t^2-4x=0,
\ee
which is exactly the envelope (\ref{s1}). Similarly, for (\ref{ex2}) we set $\varphi=\dot x$ and
we are led to the equation
\be
\varphi^2+\varphi x^2 t+x^3=0,
\ee
for which the vanishing of its discriminant gives
\be
\beta^2-4\a\g=(x^2t)^2-4x^3=x^3(xt^2-4)=0,
\ee
which, excluding the trivial solution for $x$, gives precisely the envelope (\ref{s2}).
\subsection{The case of a perfect fluid bulk}
In \cite{ack} we studied a model consisting of a single three-brane embedded in a five-dimensional
bulk space with metric
\be
\label{warpmetric}
g_{5}=a^{2}(Y)g_{4}+dY^{2},
\ee
where $g_{4}$ is the four-dimensional flat, de Sitter or anti de Sitter metric,
and an analog of perfect fluid with an energy-momentum tensor of the form
\be
\label{T old}
T_{AB}=(\rho+p) u_{A} u_{B}-p g_{AB},
\ee
where $A,B=1,2,3,4,5$ and $u_{A}=(0,0,0,0,1)$ (the 5th coordinate corresponds
to $Y$). We further assumed that the fluid satisfies a linear equation of state with
parameter $\gamma$, i.e. $p=\gamma \rho$, where the `pressure' $p$ and the `density'
$\rho$ are functions only of the fifth dimension, $Y$.

Following our set-up, the five-dimensional Einstein equations,
\be
G_{AB}=\kappa^{2}_{5}T_{AB},
\ee
can be written as
\bq
\label{syst2iii} \frac{a'^{2}}{a^{2}}&=&\frac{\kappa^{2}_{5}}{6}\rho+
\frac{k H^{2}}{a^{2}},\\
\label{syst2i}
\frac{a''}{a}&=&-\kappa^{2}_{5}\frac{(1+2\gamma)}{6}\rho,
\eq
where the prime $(\,')$ denotes differentiation with respect to $Y$ and for brevity we write
$a$ instead of $a(Y)$. The last term in (\ref{syst2iii}) is the curvature term that allows, apart from a flat
brane ($k=0$), a de Sitter ($k=1$) and anti de Sitter brane ($k=-1$) as well. Here $H$ denotes the radius of the hyperboloid representations of de Sitter (signature $-++++$) and anti-de Sitter (signature $--+++$) spaces as embedded in $\mathbb{R}^5$.
On the other hand, the equation of energy-momentum conservation,
\be
\nabla_{B}T^{AB}=0,
\ee
 becomes
\be
\label{syst2ii} \rho'+4(1+\gamma)\frac{a'}{a}\rho=0.
\ee

In \cite{ack} we used a dynamical systems method, the so-called method of asymptotic splittings,
to study, in a uniform way, both the flat and curved solutions. To apply this method we use
the variables
\be
x=a, \quad y=a', \quad w=\rho,
\ee
and write the system (\ref{syst2i}), (\ref{syst2ii}) as the following dynamical system
\bq
\label{syst2a}
x'&=&y, \\
y'&=&-(1+2\gamma)cw x,
 \\
\label{syst2c} w'&=&-4(1+\gamma)\frac{y}{x}w,
\eq
and Eq.~(\ref{syst2iii}) as
\be \label{constraint3}
\frac{y^{2}}{x^{2}}=cw+\frac{k H^{2}}{x^{2}},
\ee
where we have set $c=2A/3$ and $A=\kappa_{5}^{2}/4$.
The method of asymptotic splittings starts by identifying  all possible asymptotic behaviors of the form
\be
\label{dominant forms new} (x,y,w)=(\alpha\Upsilon^{p},\beta
\Upsilon^{q},\delta\Upsilon^{r}),
\ee
where $\Upsilon=Y-Y_{s}$, $Y_{s}$ being the position of the singularity and
\be
\label{pqm}
(p,q,r)\in\mathbb{Q}^{3}
\quad \textrm{while} \quad (\alpha,\beta,\delta)\in
\mathbb{C}^{3}\smallsetminus\{\mathbf{0}\}.
\ee
These are described in short by the dominant balances
$\mathcal{B}=\{\mathbf{a},\mathbf{p}\}$, where
$\mathbf{a}=(\alpha,\beta,\delta)$ and $\mathbf{p}=(p,q,r)$.

In \cite{ack} we found that for a flat brane the only possible
dominant asymptotic behavior around the finite-time singularity is described by the following
dominant balance
\be
\label{gB1}
_{\gamma}\mathcal{B}_{1}=\left\{\left(\alpha,\alpha
p,\frac{3}{2A}p^{2}\right),(p,p-1,-2)\right\},
p=\frac{1}{2(\gamma+1)}, \, \gamma \neq -1/2,-1.
\ee
To determine whether this balance corresponds to a particular, or, general solution
we have to calculate the eigenvalues of the matrix
\be
_{\gamma}\mathcal{K}_{1}=D\mathbf{g}(\mathbf{a})-\textrm{diag}(\mathbf{p}),
\ee
where $D\mathbf{g}(\mathbf{a})$ is the Jacobian matrix of
\be
\label{vectorfield} \mathbf{g}=\left(y,-2A\frac{(1+2\gamma)}{3}w
x,-4(1+\gamma) \frac{y}{x}w\right)^{\intercal}
\ee
and $\mathbf{a}$, $\mathbf{p}$ are determined by (\ref{gB1}). The eigenvalues of the
$_{\gamma}\mathcal{K}_{1}$ matrix constitute its spectrum, $\textrm{spec}(_{\gamma}\mathcal{K}_{1})$.
The number of non-negative eigenvalues equals 
the number of arbitrary constants that appear in the asymptotic expansions of solutions
in the form of a series defined by
\be \label{Puiseux}
\mathbf{x}=\Upsilon^{\mathbf{p}}(\mathbf{a}+
\Sigma_{j=1}^{\infty}\mathbf{c}_{j}\Upsilon^{j/s}),
\ee
where $\mathbf{x}=(x,y,w)$, $\mathbf{c}_{j}=(c_{j1},c_{j2},c_{j3})$, and
$s$ is 
the least common multiple of the denominators of
the positive eigenvalues (cf. \cite{skot}, \cite{goriely}).
The balance $_{\gamma}\mathcal{B}_{1}$ corresponds thus to the general solution in a neighborhood of the singularity in our
case if and only if it possesses two non-negative eigenvalues
(the third arbitrary constant being the position of the singularity, $Y_{s}$).
For this balance,  we found \cite{ack} that
\be
\textrm{spec}(_{\gamma}\mathcal{K}_{1})=
\left\{-1,0,\frac{1+2\gamma}{1+\gamma}\right\},
\ee
and the last eigenvalue is a function of $\gamma$,  positive when either
$\gamma<-1$, or, $\gamma>-1/2$, and negative when $-1<\gamma<-1/2$.
The cases $\gamma>-1/2$ and $\gamma<-1$ correspond to the general solution and
are both characterized by the occurrence of  finite-distance singularities which are of the
type collapse I ($a\rightarrow 0$, $a'\rightarrow \infty$, $\rho\rightarrow\infty$), and
big rip ($a\rightarrow\infty$, $a'\rightarrow -\infty$, $\rho\rightarrow\infty$)
respectively, cf. \cite{ack}.

On the other hand, for the range $-1<\gamma<-1/2$, the presence of a second negative eigenvalue leaves us with two choices:  we may either expand in
descending powers in order to meet the arbitrary constant that corresponds to the second
negative eigenvalue and find in this way the expansion of the general solution at infinity\footnote{In \cite{ack}, the solution obtained for $\gamma=-4/5$, Eqs. (3.84)-(3.86) in that paper, was wrongly termed `particular', it is a general solution of the brane equations. The correct eigenvector is the one that
corresponds to the eigenvalue $-3$ of the transpose matrix of $_{-4/5}\mathcal{K}_{1}$.
This eigenvector is $(75/(4 A\alpha),-15/(2A \alpha),1)$ and then the compatibility condition at $j=-3$
is satisfied. Therefore we do not have to set $c_{-1\,1}=0$ and the solution for
$\gamma=-4/5$ is indeed general.}, or
set this arbitrary constant equal to zero and obtain the asymptotic expansion of a particular
solution which is singular at finite distance. 
However, the particular solution obtained by
setting the arbitrary constant equal to zero, as we will now show, satisfies the equation of
the enveloping surface of our dynamical system and it is thus a special 
solution with less arbitrary constants.

Instead of looking for enveloping solutions directly from the form of the general solution, we are motivated by our considerations at the end of Section 2.1 (cf. Eqns. (2.15), (2.17)) to look for such solutions directly from the field equations. The only equation in which derivatives of the basic unknowns do not appear (much like the procedure mentioned previously) is the constraint. In addition, the constraint does not contain the independent variable (note that after eq. (2.9) and also after (2.15) we eliminated the time to express everything in terms of the parameter $c$ with respect to which we looked for envelopes) but the parameter $c$. To see this clearly, we start from the constraint of the
basic system, Eq. (\ref{constraint3}), the one-parameter family of brane `surfaces'
\be\label{pfcons}
f(x,y,w;c)=y^2-cx^2w-kH^2=0.
\ee
To consider the asymptotic structure  of our brane cosmology for small or large values of the extra dimension $\Upsilon$, we first examine whether the family of surfaces (\ref{pfcons}) has an envelope. We imagine a one-parameter family of branes parametrized by $c$ and ask whether there is an \emph{enveloping brane} to which this family asymptotes. As we show in the next section, the general solution for the case of flat branes has the same implicit form as the constraint equation (\ref{constraint3}). Since the initial conditions for the brane (Eq. (\ref{ic}) below) are determined by the arbitrary constants, it follows that they will depend on the parameter $c$ that defines the family of branes (\ref{pfcons}). It is in this indirect sense that the concept of an enveloping brane arises in the present context. We note, however, that \emph{curved} enveloping branes may also generally exist - for the problem in question they are given in the Appendix A.

For (\ref{pfcons}), since $\pa_c f=-x^2w$, the enveloping brane
is given by the following distinct
pieces:
\be\label{pfenv1}
\Sigma_1:\quad x=0,\quad y=\pm H\sqrt{k},
\ee
\be\label{pfenv2}
\Sigma_2:\quad y=\pm H\sqrt{k},\quad w=0.
\ee
The enveloping brane is thus defined as the union
\be\label{pfenv}
\Sigma=\Sigma_1\cup\Sigma_2.
\ee
We note that the envelope is not an exact solution of the field equations but an asymptotic one. Here we focus on the flat case described by $_\gamma\mathcal{B}_1$, while the envelopes of the remaining curved cases and a special flat case valid only for $\gamma=-1/2$ are completely analyzed in Appendix A. In particular, for the range $-1<\gamma<-1/2$, for example for $\gamma =-4/5$ we find \cite{ack} the dominant balance solution
\bq
\label{-4/5_B1x} x&=& \a\Upsilon^{5/2},\\
y&=& 5\a/2\Upsilon^{3/2},\\
\label{-4/5_B1w}
w&=&75/(8A)\Upsilon^{-2}.
\eq
For $\Upsilon\rightarrow 0$, we find that this asymptotic solution clearly approaches the enveloping brane since it satisfies the equation of
$\Sigma_1$ (this is seen most clearly by solving Eq. (\ref{-4/5_B1w}) for $Y$ and substituting back into the other two solutions to obtain $x,y\propto w^{-5/4},w^{-3/4}$, so that for arbitrarily large $w$ we find that asymptotically $x,y\rightarrow 0$, for the flat case we are considering here). Further,  the enveloping brane is singular. We see that the asymptotic method traced this special
solution instead of the general one, 
and we find here that this `last' enveloping brane (obtained by setting the arbitrary constant corresponding to the negative eigenvalue equal to zero) is a singular limit (it will also follow from the
next Section that for a suitable choice of the arbitrary constants, we may obtain this
solution as the envelope of the general solution that is singular).

On the other hand, the full family of solutions corresponding to the flat brane case found in \cite{ack} (i.e., Eqs. (3.84)-(3.86) in that paper) has an expansion in descending powers and is valid at infinity. This general solution blows up there in both $a,a'$ while the density $\rho$ limits to zero asymptotically as $\Upsilon\rightarrow\infty$. Transferring this solution to the finite distance position of the singularity through the transformation  $\Upsilon\rightarrow 1/\Upsilon$, we see that the corresponding solution is also singular (this movable singularity is also expected to be of the same type as that at infinity, cf. \cite{con}, chap.5).
This fact leaves no room for the avoidance of finite-distance singularities, since, as we
showed in \cite{ack}, finite singularities exist also for $\gamma<-1$ and $\gamma>-1/2$. Taking this into account,
we focus, for the rest of this paper, on re-analyzing the question of avoiding singularities
by exploiting the presence of the brane that introduces a natural symmetry in the bulk space
allowing to consider only part of it that may be non-singular.
\section{Avoidance of singularities by matching solutions}
In this section, we give the analytic solution of the dynamical system for the case of a flat
brane and examine the possibility of avoiding singularities by cutting the bulk space
and matching the solutions that are regular.

To solve the system analytically, we first substitute $k=0$ in (\ref{syst2iii})
which gives
\be
\label{syst2iiiflat}
\frac{a'^{2}}{a^{2}}=c\rho.
\ee
We then integrate (\ref{syst2ii}) to reveal the relation between $\rho$ and $a$ which
is
\be
\label{rho to a}
\rho=c_{1}a^{-4(\gamma+1)},
\ee
with $c_{1}$ an arbitrary constant. To find the solution for the warp factor $a$, we
substitute (\ref{rho to a}) in (\ref{syst2iiiflat}) and then integrate. We end up with
\be
\label{solution a}
a=\left(2(\gamma+1)\left(\pm\sqrt{\dfrac{2Ac_{1}}{3}}Y+c_{2}\right)\right)^{1/(2(\gamma+1))}, \quad \gamma\neq -1.
\ee
We now find the exact form of the fluid density by inserting the above solution
for $a$ in (\ref{rho to a}), this gives
\be
\label{solution rho}
\rho=c_{1}\left(2(\gamma+1)\left(\pm\sqrt{\dfrac{2Ac_{1}}{3}}Y+c_{2}\right)\right)^{-2}.
\ee
Substitution of our solution for $a$ and $\rho$ in (\ref{syst2i}) shows that
the latter equation is satisfied.
Our solution (\ref{solution a}) and (\ref{solution rho}) holds for all values of $\gamma$
except from $\gamma=-1$.\footnote{We may solve the system (\ref{syst2iiiflat}), (\ref{syst2i})
and (\ref{syst2ii}) for $\gamma=-1$ in order to have a complete picture of the dynamics.
Inserting $\gamma=-1$ we find that $\rho$ is a constant and $a$ has an exponential
form which means that there are no finite-distance singularities in this case. This behavior
is anticipated since $\gamma=-1$ corresponds to a cosmological constant.}
We therefore see that there exists a singularity at $\mp c_{2}\sqrt{3/(2A c_{1})}$
for all $\gamma\neq -1$. In particular, as $Y$ tends to $\mp c_{2}\sqrt{3/(2A c_{1})}$,
$\rho$ becomes divergent irrespectively of $\gamma$ ($\gamma\neq -1$). On the other
hand, the behavior of $a$ depends on $\gamma$ in the following sense: it diverges for
$\gamma<-1$ and vanishes for $\gamma>-1$.

We may apply the method of finding the enveloping surface to the general solution that
we have now. However, this is just equivalent to our study of the constraint in the previous section for the following reason. Solving  Eq. (\ref{solution a}) for  $\Upsilon$ and substituting in (\ref{solution rho}), we see that the constant $c_2$ is eliminated and we end up exactly with Eq. (\ref{rho to a}) which contains only $c_1$. Therefore the constraint (\ref{syst2iiiflat}) (or (\ref{pfcons})), which does not contain the integration constants $c_1,c_2$, can  be regarded as a  non-parametric (implicit) representation of the general solution. In this respect it plays a role similar to that of the $F=0$ equations, or Eqns. (2.15) and (2.17) of Section 2.1, with the independent variable eliminated.

From our solution (\ref{solution a})-(\ref{solution rho}), we see that it is possible to avoid
singularities, by making an appropriate choice for the range of parameters, for example
we may choose
\be
\gamma <-1 \quad \textrm{and} \quad c_{2}\leq 0,
\ee
combined with the $+$ sign for $Y<0$ and the $-$ sign for $Y>0$.
In this case, we have the solution
\be
\label{solution a_g<-1}
a=\left(2(\gamma+1)\left(-\sqrt{\dfrac{2A c_{1}}{3}}|Y|+c_{2}\right)\right)^{1/(2(\gamma+1))},
\ee
and
\be
\label{solution rho_g<-1}
\rho=c_{1}\left(2(\gamma+1)\left(-\sqrt{\dfrac{2A c_{1}}{3}}|Y|+c_{2}\right)\right)^{-2},
\ee
with the brane placed at the origin $Y=0$.
Clearly then, both $a$ and $\rho$ are non-singular since the term
$\left(-\sqrt{2A c_{1}/3}|Y|+c_{2}\right)$
is always negative. Another choice would be
\be
\gamma >-1 \quad \textrm{and} \quad c_{2}\geq 0,
\ee
with the $+$ sign for $Y>0$ and the $-$ sign for $Y<0$.
Then we would have
\be
a=\left(2(\gamma+1)\left(\sqrt{\dfrac{2A c_{1}}{3}}|Y|+c_{2}\right)\right)^{1/(2(\gamma+1))},
\ee
and
\be
\rho=c_{1}\left(2(\gamma+1)\left(\sqrt{\dfrac{2A c_{1}}{3}}|Y|+c_{2}\right)\right)^{-2}.
\ee
In the following Section, we will show that in order to obtain a finite four-dimensional Planck mass
we will have to restrict $\gamma$ in values less than $-1$. Therefore below, we analyze the
solution (\ref{solution a_g<-1})-(\ref{solution rho_g<-1}), which corresponds to $\gamma<-1$.

Naturally, we assume continuity of the warp factor and energy density.
We denote the value of an arbitrary constant $c_{i}$ at $Y>0$ ($Y<0$) by
$c_{i}^{+}$ ($c_{i}^{-}$) and find that continuity of the warp factor at $Y=0$ leads to the condition
\be
(2(\gamma+1)c_{2}^{+})^{1/(2(\gamma+1))}=(2(\gamma+1)c_{2}^{-})^{1/(2(\gamma+1))},
\ee
or, since $c_{2}^{+}$ and $c_{2}^{-}$ are real numbers, we have
\be
\label{c2}
c_{2}^{+}=\pm c_{2}^{-},
\ee
depending on the value of $\gamma$.
Similarly, continuity of the density gives
\be
\dfrac{c_{1}^{+}}{(c_{2}^{+})^{2}}=\dfrac{c_{1}^{-}}{(c_{2}^{-})^{2}},
\ee
and using (\ref{c2}) we find
\be
\label{c1}
c_{1}^{+}=c_{1}^{-}.
\ee
On the other hand, the jump of the extrinsic curvature
$K_{\alpha\beta}=1/2(\partial g_{\alpha\beta}/\partial Y)$ ($\alpha,\beta=1,2,3,4$), is given by
\be
\label{junction}
K_{\alpha\beta}^{+}-K_{\alpha\beta}^{-}=-\kappa_{5}^{2}\left(S_{\alpha\beta}-
\dfrac{1}{3}g_{\alpha\beta}S \right),
\ee
where the surface energy-momentum tensor $S_{\alpha\beta}$ (defined only on the brane
and vanishing off the brane) is taken to be
\be
\label{surface tensor}
S_{\alpha\beta}=-g_{\alpha\beta}f(\rho),
\ee
with $f(\rho)$ denoting the brane tension and $S=g^{\alpha\beta}S_{\alpha\beta}$ the
trace of $S_{\alpha\beta}$. Substitution of (\ref{solution a_g<-1}) and (\ref{surface tensor})
in (\ref{junction}), leads to a junction condition for the arbitrary constants
\be
\label{j1}
\sqrt{c_{1}}\left(\dfrac{1}{c_{2}^{+}}+\dfrac{1}{c_{2}^{-}}\right)=
 4\sqrt{\dfrac{2 A}{3}}(\gamma+1)f(\rho(0)),
 \ee
from which we see that we have to choose the plus sign in (\ref{c2}) and then (\ref{j1})
becomes
\be\label{ic}
\dfrac{\sqrt{c_{1}}}{c_{2}}=2\sqrt{\dfrac{2 A}{3}}(\gamma+1)f(\rho(0)).
\ee
\subsection{Planck Mass}
Another condition we need to check is whether the solution we have found for the warp factor, $a$, leads to a
finite four-dimensional Planck mass, for some range of the parameter $\gamma$. The
value of the four-dimensional Planck mass, $M_{p}^{2}=8\pi/\kappa$, is determined by
the following integral \cite{forste}
\be
\frac{\kappa_{5}^{2}}{\kappa}=\int_{-Y_{c}}^{Y_{c}}a^{2}(Y)dY.
\ee
For our solution, Eq.~(\ref{solution a_g<-1}), the above integral becomes,
\beq
& &\int_{-Y_{c}}^{Y_{c}}
\left(2(\gamma+1)\left(-\sqrt{\dfrac{2 A c_{1}}{3}}|Y|+c_{2}\right)\right)^{1/(\gamma+1)}dY
=\\
&=&\dfrac{1}{2(\gamma+2)}\sqrt{\dfrac{3}{2A c_{1}}}\left(2(\gamma+1)\left(\sqrt{\dfrac{2Ac_{1}}{3}}Y
+c_{2}\right)\right)^{(\gamma+2)/(\gamma+1)}|_{-Y_{c}}^{0}-\\
&-&\dfrac{1}{2(\gamma+2)}\sqrt{\dfrac{3}{2A c_{1}}}\left(2(\gamma+1)\left(-\sqrt{\dfrac{2Ac_{1}}{3}}Y
+c_{2}\right)\right)^{(\gamma+2)/(\gamma+1)}|_{0}^{Y_{c}},
\eeq
In the limit $Y_{c}\rightarrow\infty$, the Planck mass remains finite only for
\be
\label{finite_planck}
-2<\gamma<-1,
\ee
and takes the form
\be
\frac{\kappa_{5}^{2}}{\kappa}=\sqrt{\dfrac{3}{2A c_{1}}}
\dfrac{(2(\gamma+1)c_{2})^{\frac{\gamma+2}{\gamma+1}}}{
\gamma+2}.
\ee
This result agrees with the analysis of \cite{forste} which used a particular field theory model
involving a scalar field with non-trivial kinetic terms.
\subsection{Energy conditions}
In the previous Section, we showed that the requirement for a finite four-dimensional Planck mass restricts
$\gamma$ in the interval $(-2,-1)$. Here we discuss whether this range of $\gamma$
is allowed, or, somehow prohibited by physical constraints such as the energy positivity
conditions.
To answer this question, we first construct the weak and strong energy conditions for our
type of matter (\ref{T old}) and then examine for which ranges of $\gamma$ they hold true.

We note that our metric (\ref{warpmetric}) and our fluid are static with respect to $t$.
We may reinterpret our fluid as an anisotropic one having the following energy momentum
tensor
\be
\label{T new}
T_{AB}^{\textrm{new}}= (\rho^{\textrm{new}}+
p^{\textrm{new}})u_{A}^{new}u_{B}^{\textrm{new}}
+p^{\textrm{new}}g_{\alpha\beta}\delta_{A}^{\alpha}\delta_{B}^{\beta}+
p_{Y}g_{55}\delta_{A}^{5}\delta_{B}^{5},
\ee
where $u_{A}^{\textrm{new}}=(a(Y),0,0,0,0)$, $A,B=1,2,3,4,5$ and $\alpha,\beta=1,2,3,4$.
When we combine (\ref{T old}) with (\ref{T new}) we get the following set of relations
\bq
\label{p y to rho}
p_{Y}&=&\rho\\
\label{rho new}
\rho^{\textrm{new}}&=&p\\
\label{p new}
p^{\textrm{new}}&=&-p.
\eq
The last two relations imply that
\be
\label{p new to rho new}
p^{\textrm{new}}=-\rho^{\textrm{new}},
\ee
which means that this type of matter satisfies a cosmological constant-like equation of state.
Imposing further $p=\gamma\rho$ and using (\ref{p y to rho}) leads to
\be
\label{py to p}
p_{Y}=\frac{p}{\gamma}.
\ee
Substituting (\ref{p new}), (\ref{p new to rho new}) and (\ref{py to p}) in (\ref{T new}), we
find that
\be
T_{AB}^{\textrm{new}}= -p g_{\alpha\beta}\delta_{A}^{\alpha}\delta_{B}^{\beta}+
\frac{p}{\gamma}g_{55}\delta_{A}^{5}\delta_{B}^{5}.
\ee

We are now ready to form the energy conditions for our type of matter.
We begin with the weak energy condition according to which, every future-directed
timelike vector $v^{A}$ should satisfy
\be
T_{AB}v^{A}v^{B}\geq 0.
\ee
This condition implies that the energy density should be non negative for all forms of
physical matter \cite{wald}. Here we find (see Appendix B, Subsection B.1) that it
translates to
\be
\label{wec_p}
p\geq 0 
\ee
and
\be
\gamma>0, \quad \textrm{or},\quad  \gamma<-1.
\ee
On the other hand, we have to exclude the case $\gamma<-1$ since when it is combined with (\ref{wec_p}) it gives a
negative $\rho$ which is in contradiction with (\ref{syst2iiiflat}). We end up with the condition
\be
\label{wec}
p\geq 0 \quad \textrm{and}\quad  \gamma>0.
\ee

The strong energy condition, on the other hand, demands that
\be
\left(T_{AB}-\dfrac{1}{3}T g_{AB}\right)v^{A}v^{B}\geq 0,
\ee
for every future-directed timelike vector $v^{A}$. In our case,
this condition leads to the following restrictions for $p$ and $\gamma$
(see Subsection B.2): We should either have $p=0$, or,
\be
\label{sec_1}
p< 0 \quad \textrm{and}\quad -1\leq\gamma<0,
\ee
or,
\be
\label{sec_2}
p> 0 \quad \textrm{and}\quad 0<\gamma\leq 1.
\ee

The conditions (\ref{wec}) and (\ref{sec_1}), (\ref{sec_2}) show that the weak and strong
energy  conditions restrict $\gamma$ to be, in either case, greater than $-1$. This means
that the type of matter considered in this paper, cannot at the same time satisfy
the energy conditions \emph{and} the requirement for a finite Planck mass given by
(\ref{finite_planck}).
\section{Conclusions}
In the first part of this paper, we studied the effect that the existence of envelopes brings into the
dynamics of the cosmological model of \cite{ack}, consisting of a three-brane embedded in a
five-dimensional fluid bulk satisfying an analog of an equation of state
$p=\gamma\rho$. The fluid leads to the appearance of singularities
within finite distance for $\gamma<-1$ and $\gamma>-1/2$, \cite{ack}. For $-1<\gamma<-1/2$,
we showed presently that singular behavior is dictated by both a particular solution
that cannot be assumed as a less general behavior since it belongs to the singular part of the
enveloping surface of the dynamical system in question (enveloping brane), and also by the fact that the general asymptotic solution at infinity when transformed to the finite distance position is also singular.

The fact that the general solution as well as the enveloping branes are both singular for all values of $\gamma$
once the whole indivisible bulk is considered,  led us to the second part of this paper,
in which we examined the possibility of avoiding the singularities by cutting the bulk space and matching
the solutions that correspond to its regular part at the position of the brane. The regular solutions that we obtain in
this way, give a finite four-dimensional Planck mass for $-2<\gamma<-1$.
The next question is if this range can be realized by a physical system. For this we examined
an analog of the weak and strong energy conditions for the five-dimensional fluid and found
that the above region of $\gamma$ does not satisfy them.
This seems to be consistent with the result of \cite{forste} that used a field theory model
realizing such equation of state and found a tachyonic instability.

It is in principle possible that a singular general solution may possess an envelope having a regular  piece. Unfortunately, as we showed in this work, this is not the case with the family of models considered in the present paper.
It is an open question
whether the form we used for the weak and strong energy conditions for the bulk system
can be avoided (for example, by considering other canonical reductions), or whether a more  general equation of state can be realized by considering
for instance $\gamma$ non constant but a function of the fifth coordinate $Y$. This would probably require analyzing a sort of interacting mixture on the bulk.
\section*{Acknowledgements}
Work supported in part by the European Commission under the ERC
Advanced Grant 226371.
I.K. is grateful to CERN-TH, where part of her work was done, for financial support that
made her visits there possible and for allowing her to use its excellent facilities. We thank an anonymous referee whose very useful comments and suggestions helped to produce a much clearer version of this work.
\appendix
\section{Appendix: Envelopes}
In this Section, we list the dominant balances that we found for the case of a perfect fluid
bulk in \cite{ack} and find the part of the enveloping surface to which they belong, that is we give the explicit structure of the enveloping branes.

The dominant balances are
\bq
_{\gamma}\mathcal{B}_{1}&=&
\left\{\left(\alpha,\alpha
p,\frac{3}{2A}p^{2}\right),(p,p-1,-2)\right\},
p=\frac{1}{2(\gamma+1)}, \, \gamma \neq -1/2,-1,\label{preveq}\\
_{\gamma}\mathcal{B}_{2}&=&\{(\alpha,\alpha,0),(1,0,-2)\},\quad \gamma \neq -1/2,\\
_{-1/2}\mathcal{B}_{3}&=&\{(\alpha,\alpha,0),(1,0,r)\},
\\
_{-1/2}\mathcal{B}_{4}
&=&\{(\alpha,\alpha,\delta),(1,0,-2)\},\\
_{-1/2}\mathcal{B}_{5}&=&\{(\a,0,0), (0,-1,r)\},\label{Bfive} \eq
where $_{-1/2}\mathcal{B}_{i}\equiv_{\gamma=-1/2}\mathcal{B}_{i}$.

The first balance, $_{\gamma}\mathcal{B}_{1}$, satisfies the constraint equation,  Eq.~(\ref{constraint3}), for $k=0$ immediately, that is in the first step of the method of asymptotic splittings, where $j=0$ in Eq.~(\ref{Puiseux}). This does not imply, however, that it cannot also describe a solution for a curved brane. In fact, for this balance the curvature term, $k H^{2}/a^{2}$, has a subleading behavior that can contribute later on in the series expansion, in the way that it sets one of the arbitrary constants equal to a specific value. For example, for $\gamma =0$ the curvature term contributes at the second step of the asymptotic method, that is at $j=1$, and sets the arbitrary constant $c_{1\,3}$ equal to $-3kH^{2}/(4A \a ^{2})$. We therefore see that for $k=0$, the arbitrary constant $c_{1\,3}$ is vanishing which means that the balance $_{\gamma}\mathcal{B}_{1}$ constitutes the general solution of the system for a flat brane. For a curved brane, on the other hand, one needs to go further up the series expansion (one more step) to find the asymptotic form of the general solution. A similar behavior is found for the balance $_{-1/2}\mathcal{B}_{5}$. In this latter case, the curvature term sets the arbitrary constant $c_{11}^{2}$ equal to $k H^{2}$ at $j=2$.
The next two balances, $_{\gamma}\mathcal{B}_{2}$ and $_{-1/2}\mathcal{B}_{3}$, satisfy the constraint equation only when $\a^{2}=kH^{2}$ and describe therefore solutions of curved branes. Finally, for the balance $_{-1/2}\mathcal{B}_{4}$, we find that it describes a solution for a curved or flat brane with $\d=(3/(2A))(1-kH^{2}/\a^{2})$.

The second balance, $_\g\mathcal{B}_2$, for $\g<-1/2$, for example $\gamma=-3/4$, gives
\bq
\label{-3/4_B2x}
x&=& \a\Upsilon+\frac{A\a}{6}c_{1\,3}\Upsilon^{2}+\cdots,\\
y&=& \a+\frac{A\a}{3} c_{1\,3}\Upsilon+\cdots,\\
\label{-3/4_B2w} w&=&  c_{1\,3}\Upsilon^{-1}+\cdots, \eq
where $c_{1\,3}$ is an arbitrary constant such that $c_{1\,3}\neq 0$.
This is on $\Sigma_1$, Eq. (\ref{pfenv1}), since as $\Upsilon\rightarrow 0$,
it approaches this part of the envelope.
The same balance for $\gamma>-1/2$, for example $\gamma=0$,
gives
\bq
\label{0_B2x}
x&=& \a\Upsilon+c_{-1\,1}-A\a/3c_{-2\,3}\Upsilon^{-1}+\cdots,\\
y&=& \a+A\a/3 c_{-2\,3}\Upsilon^{-2}+\cdots,\\
\label{0_B2w} w&=&  c_{-2\,3}\Upsilon^{-4}+\cdots,
\eq
where $c_{-1\,1}$ and $c_{-2\,3}$ are arbitrary constants. For $\Upsilon\rightarrow \infty$
we see that this is on $\Sigma_2$.

The balance $_{-1/2}\mathcal{B}_3$ for $r=-3$ 
implies the following asymptotic behavior
 \bq \label{-1/2_B_3x}
x&=& \a\Upsilon+\cdots,\\
y&=& \a+\cdots,\\
\label{-1/2_B_3w}
w&=& c_{1\,3}\Upsilon^{-2}+\cdots,
\eq
with $c_{1\,3}$ an arbitrary constant. Taking $\Upsilon\rightarrow 0$, shows that this
is on $\Sigma_1$.
For the same balance but for $r=0$  we have
\bq
\label{-1/2_B3x}
x&=& \a\Upsilon+c_{-1\,1},\\
y&=& \a,\\
\label{-1/2_B3w} w&=&  c_{-2\,3}\Upsilon^{-2}+\cdots, \eq
where $c_{-1\,1} $ and $c_{-2\,3}$ are arbitrary constants.
Taking $\Upsilon\rightarrow \infty$ demonstrates that this is on $\Sigma_2$.

Lastly, for the balance $_{-1/2}\mathcal{B}_5$ and necessarily for 
$r=1$ (see \cite{ack}) we find
\bq
\label{-1/2_B4xn}
x&=& \a+c_{1\,1}\Upsilon+\cdots,\\
\label{-1/2_B4yn}
y&=& c_{1\,1}+\cdots,\\
\label{-1/2_B4wn} w&=&  0+\cdots.
\eq
Taking $\Upsilon\rightarrow 0$, we see that this asymptotic solution belongs to the
piece $\Sigma_2$, Eq. (\ref{pfenv2}) of the enveloping surface for both flat and curved branes since
the arbitrary constant $c_{1\,1}$ in Eq. (\ref{-1/2_B4yn}) is set to the value $c_{11}^{2}=kH^{2}$ by the constraint equation, as mentioned in the beginning of this Section.

\section{Appendix: Energy conditions}
In this Appendix, we form the weak and strong energy conditions for the type of fluid
considered in this paper by adopting the formalism
expounded in \cite{poisson} p. 28-31 for our model. Since our medium is clearly anisotropic, it is not possible to make  statements similar to the situation in general relativity.   We assume that the energy-momentum
tensor given by Eq. (\ref{T new}),  can be decomposed in the following way,
\be
\label{T_decomp}
\hat{T}^{AB}=\rho^{\textrm{new}}\hat{e_{1}}^{A}\hat{e_{1}}^{B}+
p^{\textrm{new}}\hat{e_{2}}^{A}\hat{e_{2}}^{B}+
p^{\textrm{new}}\hat{e_{3}}^{A}\hat{e_{3}}^{B}+
p^{\textrm{new}}\hat{e_{4}}^{A}\hat{e_{4}}^{B}+
p_{Y}\hat{e_{5}}^{A}\hat{e_{5}}^{B},
\ee
where the vectors $\hat{e_{M}}^{A}$ consist
an orthonormal basis and
\be
g_{AB}\hat{e_{M}}^{A}\hat{e_{N}}^{B}=\eta_{M N},
\ee
$\eta_{M N}$ being the Minkowski metric. Inserting the expressions of $p_{Y}$,
$\rho^{\textrm{new}}$ and $p^{\textrm{new}}$, given by
Eqs.~(\ref{p y to rho})-(\ref{p new}), in (\ref{T_decomp}) we find,
\be
\label{T decomp2}
\hat{T}^{AB}=p\hat{e_{1}}^{A}\hat{e_{1}}^{B}-
p\hat{e_{2}}^{A}\hat{e_{2}}^{B}-
p\hat{e_{3}}^{A}\hat{e_{3}}^{B}-
p\hat{e_{4}}^{A}\hat{e_{4}}^{B}+
\dfrac{p}{\gamma}\hat{e_{5}}^{A}\hat{e_{5}}^{B}
\ee
To set up the energy conditions we will need a normalized future-directed timelike vector
$v^{A}$ which may be decomposed in the following way
\be
\label{v A}
v^{A}=w(\hat{e_{1}}^{A}+a \hat{e_{2}}^{A}+b \hat{e_{3}}^{A}+c \hat{e_{4}}^{A}+d \hat{e_{5}}^{A}),
\quad w=(1-a^{2}-b^{2}-c^{2}-d^{2})^{-1/2},
\ee
where $a,b,c,d$ are such that
\be
\label{v timelike}
a^{2}+b^{2}+c^{2}+d^{2}<1.
\ee
In the following, we derive the energy conditions the final forms of which were
used in Subsection 3.2.
\subsection{Weak energy condition}
For any timelike vector $t^A$, we assume that $T_{AB}t^At^B\geq 0$, and so  the energy density measured by an observer
having velocity $v^{A}$, that is represented by the quantity $T_{AB}v^{A}v^{B}$,
must be non-negative,
i.e.,
\be
T_{AB}v^{A}v^{B}\geq 0.
\ee
Substituting $\hat{T}_{AB}$ and $v^{A}$ in the above relation we get
\be
p-p a^{2}-p b^{2}-p c^{2}+\dfrac{p}{\gamma}d^{2}\geq 0.
\ee
Since $a,b,c,d$ are arbitrary, we may choose $a=b=c=d=0$ (and similarly $a=c=d=0$, or,
$a=b=d=0$) and find that
\be
\label{wec p}
p\geq 0.
\ee
On the other hand, setting $a=b=c=0$ gives
\be
p+\dfrac{p}{\gamma}d^{2}\geq 0.
\ee
Because of (\ref{v timelike}), we must have $d^{2}<1$ which leads to
\be
\dfrac{\gamma+1}{\gamma}p>0.
\ee
In combination with (\ref{wec p}), we find two possible ranges for $\gamma$:
\be
\gamma>0, \quad \textrm{or},\quad \gamma<-1.
\ee
The second range of $\gamma$ is excluded, since it implies
\be
\rho=\dfrac{p}{\gamma}<0
\ee
which contradicts Eq. (\ref{syst2iiiflat}). Therefore, the weak energy condition restricts $p$
and $\gamma$ in the following way,
\be
p\geq 0 \quad \textrm{and} \quad \gamma>0.
\ee
\subsection{Strong Energy condition}
The strong energy condition states that
\be
\label{sec}
\left(T_{AB}-\dfrac{1}{3}T g_{AB}\right)v^{A}v^{B}\geq 0,
\ee
where $T$ is the trace of $T_{AB}$. In our case,
\be
\label{trace T}
\hat{T}=-4p+\dfrac{p}{\gamma}.
\ee
Inserting in the general form of the strong energy condition given by (\ref{sec}), the
Eqs.~(\ref{v A}), (\ref{trace T}) and $\hat{T}_{A B}$ we find,
\be
w^{2}\left(p-p a^{2}-p b^{2}-p c^{2} +\dfrac{p}{\gamma}d^{2}\right)\geq
-\dfrac{1}{3}\left(-4p+\dfrac{p}{\gamma}\right).
\ee
For $a=b=c=d=0$, Eq.~(\ref{v A}) gives $w=1$ and the above relation becomes
\be
\label{sec 1}
\dfrac{\gamma-1}{\gamma} p\leq 0.
\ee
Putting in turn, $a=c=d=0$, $a=b=d=0$, or, $b=c=d=0$, leads to the same result.
On the other hand, taking $a=b=c=0$,  gives
\be
\dfrac{1}{1-d^{2}}\left(p+\dfrac{p}{\gamma}d^{2}\right)\geq
\dfrac{1}{3}\left(4p-\dfrac{p}{\gamma}\right),
\ee
or,
\be
-p+\dfrac{p}{\gamma}\geq -d^{2}\left(4p+2\dfrac{p}{\gamma}\right)
\ee
and using the fact that $d^{2}<1$, we end up wth
\be
\label{sec 2}
\dfrac{\gamma+1}{\gamma}p\geq 0.
\ee
Combining Eqs.~(\ref{sec 1}) and (\ref{sec 2}) we see that we have either $p=0$ or
\bq
\label{wec i}
p< 0 \quad &\textrm{and}& \quad -1\leq\gamma<0,\quad\textrm{or},\\
p> 0 \quad &\textrm{and}& \quad 0<\gamma\leq 1.
\eq

\end{document}